\documentclass{ws-p8-50x6-00}

\begin{document}

\title{Bose-Einstein correlations in cascade processes and non-extensive
statistics}

\author{O.V.Utyuzh and G.Wilk}

\address{The Andrzej So\l tan Institute for Nuclear Studies; 
Ho\.za 69; 00-689 Warsaw, Poland\\E-mail: utyuzh@fuw.edu.pl and
wilk@fuw.edu.pl}

\author{Z.W\L odarczyk}

\address{Institute of Physics, Pedagogical University;
Konopnickiej 15; 25-405 Kielce, Poland\\
E-mail: wlod@pu.kielce.pl}


\maketitle

\abstracts{
We discuss the effect of nonextensivity of the emitting source on
the Bose-Einstein correlations (BEC). This is done numerically by
comparing cascade hadronization model (CAS), which is known to exhibit
fractal structure in both space-time and phase-space, with its
equivalent obtained from the information theory approach (MaxEnt),
in which hadronization proceeds uniformly in the phase-space. To this
end we have developed a new method of accounting for BEC in Monte
Carlo event generators, which preserves all kinematics of the
hadronization process.} 

Some time ago problem of sensitivity of Bose-Einstein correlations
(BEC) \cite{BEC} to the possible fractal structure of emitting
hadronic source has been formulated \cite{B}. Recently we have
addressed this problem numerically by modelling such source in terms
of the cascade process taking place both in phase-space and in
space-time \cite{CAS1,CAS2}. In addition to the expected feature of
intermittency observed in its multiparticle distributions it also
shows a kind of modified L\'evy distribution in space-time variables,
which can be interpreted as a signal of nonextensivity arising due to
the space-time fractal structure of such source \cite{T}. In this
context it is worth to note recent investigations of the possible
continuous emission from the otherwise hydrodynamically described
expanding hadronic source, which leads also to a kind of foam-like
(multifractal) space-time structure noticeably influencing the BEC
\cite{HYDRO}. As we have  demonstrated \cite{CAS2} BEC are, indeed,
sensitive to the nonextensivity parameter $q$ which influences
spacio-temporal development of the cascade process, i.e., its
fractality. They are also sensitive to the particular ("slow" or
"fast") way in which cascade develops (quantified by the mean
life-time of the cascade link parameter $\tau$ \cite{CAS1,CAS2}).
However, in \cite{CAS2} we did not compare explicitely nonextensive
(such as cascade) approach to hadronization with an extensive one,
our analysis concerned properties of the cascade process itself. Such
comparison will be performed here. However, the "afterburner" method
of incorporating BEC into event generators \cite{AFTER} used in
\cite{CAS1,CAS2} is not satisfactory for this purpose because it changes
the original event, making comparison of events coming from different
hadronization models very difficult. The more satisfactory weighting
procedures \cite{WEIGHTS,WIBIG} are, on the other hand, too much
complicated for our purposes. The same is true for procedures relying
on shifting of momenta \cite{SHIFT}, which in addition also change the
initial energy-momentum balance.

We shall propose instead a new, simple procedure of modelling BEC,
which preserves both the energy-momenta of all produced secondaries
and their total multiplicity distributions (as well as their 
intermittency pattern). It changes, however, the charge allocation
(if any) of the produced secondaries, preserving at the same time
both the total charge of the initial source and multiplicities of
charged and neutral secondaries resulting from a given event
generator. Suppose that in the $l^{th}$ event ($l=1,\dots,N_{event}$)
our generator provides us with $n_l = n_l^{(+)} + n_l^{(-)} +
n_l^{(0)}$ particles. Keeping their energy-momenta and
spacio-temporal positions intact we shall now allocate to them anew
the charges and this will be done in the following way:  
\begin{itemize}
\item[{\bf 1.}] One chooses randomly, with weights proportional to
$p^{(+)}_l=n_l^{(+)}/n_l$, $p^{(-)}_l=n_l^{(-)}/n_l$
and $p^{(0)}_l=n_l^{(0)}/n_l$, the {\it SIGN} (from: "+", "-"
or "0") and attaches it to the particle $(i)$ chosen randomly from
particles produced in this event and not yet reassigned new  charges.
\item[{\bf 2.}] One calculates distances in momenta, $\delta_{ij}(p) =
\left|p_i - p_j\right|$, between the chosen particle $(i)$ and all
other particles still without signs and arranges them in order of
the ascending $\delta_{ij}(p)$ with $j=1$ denoting the nearest
neighbour of particle $(i)$. To each $\delta_{ij}(p)$ an appropriate
weight $P(i,j)$ is then assigned (the form of which will be discussed
below). 
\item[{\bf 3.}] One selects from a uniform distribution a random
number $r \in (0,1)$. If $n_{SIGN} > 0$, i.e., if there are still 
particles of given {\it SIGN} with not reassigned charges, 
one checks the previously
selected particles in ascending order of $j$ and if $r < P(i,j)$ 
then charge {\it SIGN} is assigned also to the
particle $(j)$, the multiplicity of particles with this {\it SIGN} is
reduced by one, $n_{SIGN} = n_{SIGN} -1$, and next particle, $j=j+1$,
is selected from that bunch. If the new $n_{SIGN} = 0$ one returns to
point $(1)$ but with the updated values of probabilities
$p_l^{(+)},~p_l^{(-)}$ and $_l^{(0)}$. 
However, if  $r> P(i,j)$ then one returns to $(1)$, again with the updated
values of $p_l^{(+)},~p_l^{(-)}$ and $p_l^{(0)}$.
Procedure finishes when $n_+ = n_- = n_0 = 0$, in which case  one
proceeds to the next event.
\end{itemize}
It is important to realize that the above method of choice of
particles of the same {\it SIGN} leads to a geometrical
(Bose-Einstein) distribution of particles in this group (cell) (for
$P(ij)=P=$const its mean multiplicity equals $P/(1-P)$, in fact
because it contains also $n=0$, therefore in the algorithm it will be
greater by $1$). In this way one accounts for the bosonic character
(Bose-Einstein statistics) of the produced particles and after
application of this procedure they show strong tendency to occupy the
same cell in the phase-space (defined, for example as wave packet in
the momemtum space centered on the mean momentum of selected
particles)\cite{BSWW}. That this will show up in the $2-$particle BEC
correlation function $C_2(Q)=\frac{\sigma\left(p_1,p_2\right)}
{\sigma\left(p_1\right)\sigma\left(p_2\right)} > 1$ for $Q
=|p_1-p_2|\rightarrow 0$ was already demonstrated in statistical
model based on information theory (with conservation of charges
imposed) \cite{OMT}. One of the main parameters of this model was the
size of phase-space cells containing particles of the same charge,
which was therefore fixed and the same for all events. In our case
both the sizes of phase-space cells and their number depend crucially
on the weight parameters $P(ij)$ and can therefore vary both from
event to event and also in a given event. Notice that, because we do
not limit {\it a priori} the number of particles which can be put
into given cell, we are, in fact, getting in this way automatically BEC
of {\it all orders}. It means that $C_2$ presented in Figs. 1 and 2
are calculated in the environmnet of multiparticle BEC and as
such can exceed $2$ at some circumstances (as demonstrated in
\cite{WIBIG}). 

It is instructive to realize what kind of dynamical picture this
method corresponds to in the case of cascade model. Notice that we do not change
initial energy-momentum flow here, however, we do profoundly change
the initial charge flow resulting from our event generator. Taking
any example of cascade described in \cite{CAS1}, allocating to final
secondaries charges according to the proposed procedure and working
then out charge flow backwards, one encounters strong charge fluctuations
with multiple charges occuring in branching vertices, which were
not present there originally (at the same time total charge of the
whole system is at every cascade step always equal to the initial
charge of the mass $M$ initiating cascade). This observation is, in
fact, a general one: precisely the allowance for such charge
fluctuations leads to the occurence of like-charge bunching,
which in turn, are interpreted as effect of BEC. In our case one
could argue that such multicharged vertices should be introduced into
the scheme of the hadronization cascade process itself (i.e., already
into our Monte Carlo event generator). This is, however, an
impossible task because it leads to unsurmountable problems with
their subsequent proper deexcitation to single charged final
particles. The relaxation of control over the initial charge flow (or
lack of such altogether) is therefore necessary condition of
applicability of the proposed algorithm. 

As in all other approaches it will be important what kind of weight
factors $P(ij)$ we shall choose. Two such choices will be
demonstrated here: $P(ij)=$ const $= 0.5$ and $P(ij) =
\exp[- \frac{1}{2}\delta^2_{ij}(x)\cdot \delta^2_{ij}(p)]$. This
later form uses the available information on the particle production
provided by event generator. One can argue that this is, in a sense,
a "the most natural form" because of the following: if particles
$(ij)$ would be described by the wave packets in the space-time,
their widths would follow momentum separation $\delta_{ij}(p)$ and the
corresponding probability distribution in $\delta_{ij}(x)$ would be
of the above gaussian form. 

\vspace{-2mm}
{ 
\begin{figure}[ht]
\begin{minipage}[ht]{5.75cm}
\epsfxsize=13pc 
\centerline{\epsfbox{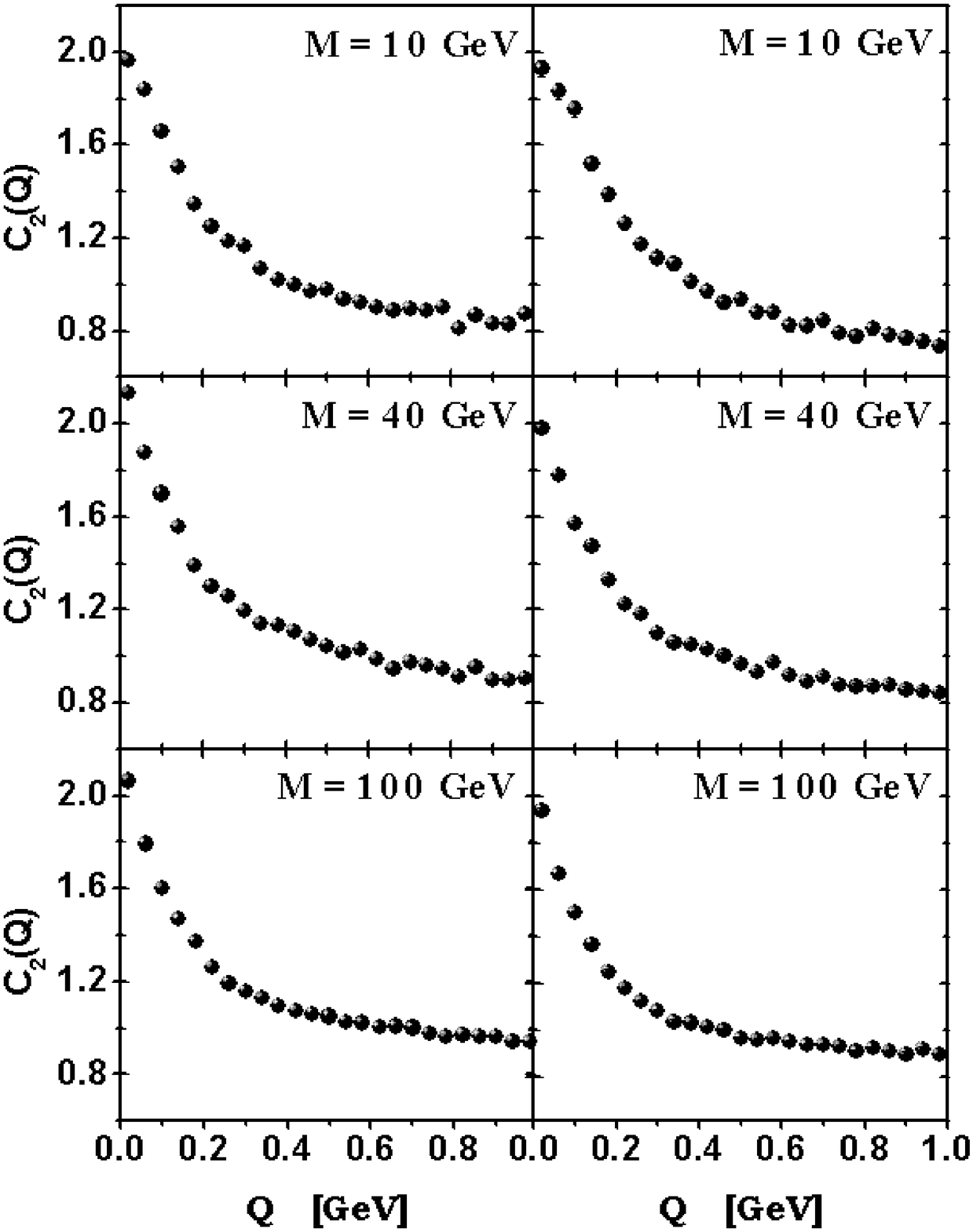}}
\vspace{-2mm}
\caption{Comparison of BEC for CAS (left panels) and MaxEnt (right
panels) types of the emmiting one-dimensional sources of masses
$M=100$, $40$ and $10$ GeV for constant value of parameter $P=0.5$
(see text for details). \label{fig:1}}
\end{minipage}
\hfill
\begin{minipage}[ht]{5.75cm}
\epsfxsize=13pc 
\centerline{\epsfbox{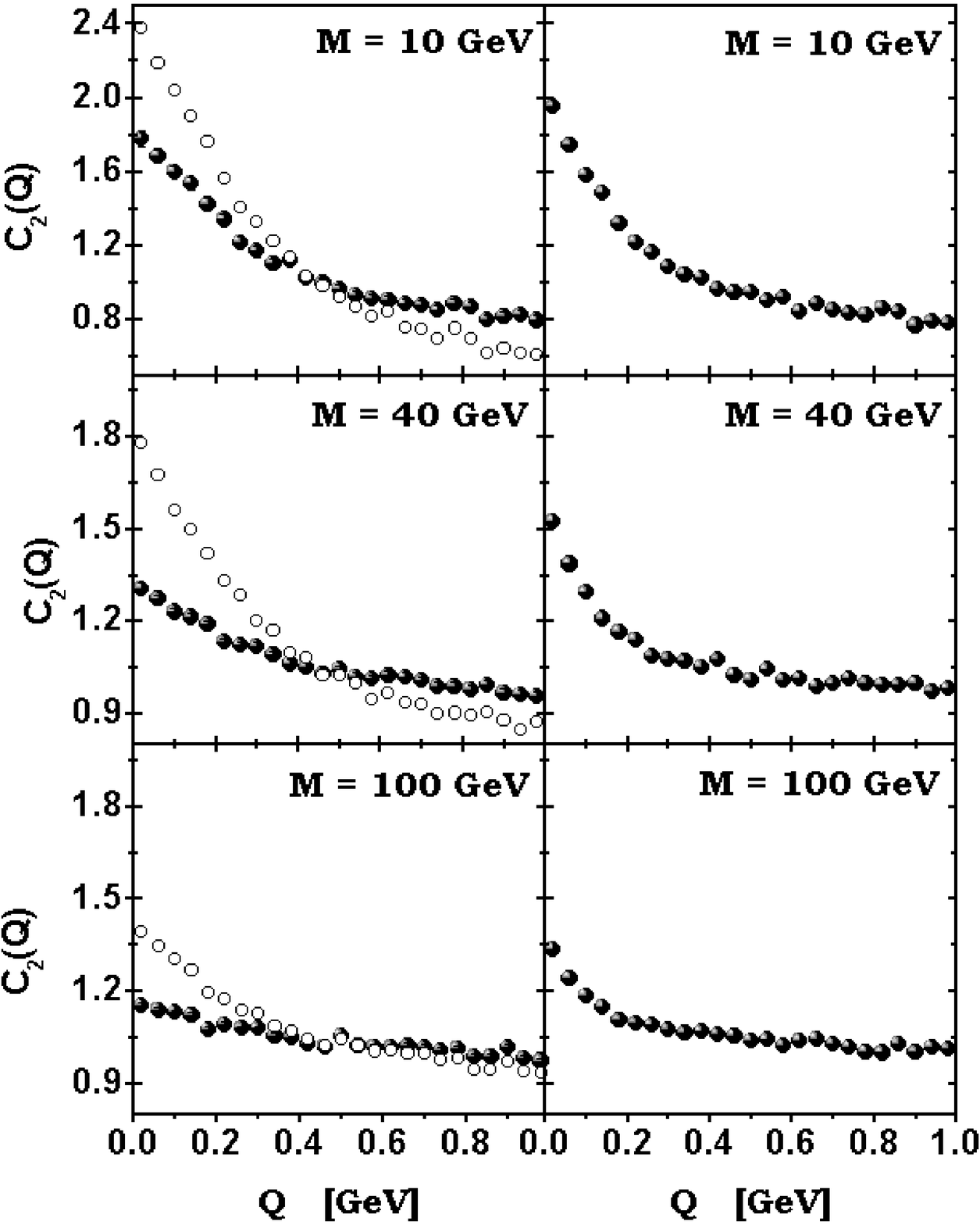}} 
\vspace{-2mm}
\caption{The same as in Fig. (\ref{fig:1}) but for the "most natural"
choice of $P$ and for two different types of cascade evolution 
characterized by constant $\tau=0.2$ fm - full symbols - and 
mass-dependent $\tau = 1/M$ -  open symbols  - (see text for details).
\label{fig:2}}
\end{minipage}
\end{figure}
}
\vspace{-2mm}

With this algorithm we can now easily compare the truly fractal source
of hadronization provided by the cascade (CAS) with the most simple
one corresponding to the instantenous hadronization in the whole
available phase space, as - for example - provided by the
maximalization of the information entropy approach (MaxEnt)
\cite{EIWW}. This is done in the following way: to each CAS event
characterized by the multiplicity $n_l$ one builds the corresponding
MaxEnt event according to the procedure outlined in \cite{EIWW} (i.e.,
one calculates the corresponding Lagrange multiplier $\beta_l$ or
"temperature" $T_l=1/\beta_l$). Using now the same
multiplicities, $n_l,~n^{(+)}_l,~n^{(-)}_l$ and $n^{(0)}_l$, as in CAS
one calculates the corresponding BEC. The results for constant
weight $P$ are given in Fig. 1 whereas Fig. 2 contains results using
the most natural choice of the weights $P(ij)$. In the case on MaxEnt
we argue that it is given by the following form: $P(ij)\, =\,
\exp\left[-\, \frac{\left( p_i - p_j\right)^2}{2k\mu_T T_l}\right].$
Here $p_{i,j}$ are the momenta of the particles considered, $\mu_T$
denotes their transverse mass, $k$ is Boltzmann constant and $T_l$ is
the mentioned above "temperature" of the $l^{th}$ event. 

\vspace{-2mm}
\begin{table}[ht]
\caption{List of parameters $\gamma$, $\lambda$ and $R$ (in $fm$) fitting 
data shown in Figs. 1 and 2 by the following formula:
$C_2(Q) = \gamma\cdot [1 + \lambda \exp (-R\cdot Q)]$. 
\label{tab:exp}}
\begin{center}
\footnotesize
\begin{tabular}{|c|c|c|c|c|c|c|c|}
\hline
 & $M$ &\multicolumn{2}{|c|}{$10$ GeV}&
        \multicolumn{2}{|c|}{$40$ GeV}&
        \multicolumn{2}{|c|}{$100$ GeV}\\
\hline
\multicolumn{2}{|l|}{Model used:} & ~~CAS~~ & MaxEnt& ~~CAS~~ & MaxEnt & ~~CAS~~ & MaxEnt \\
\hline
       & $\gamma$  & 0.84 & 0.72 & 0.92 & 0.81 & 0.97 & 0.88 \\ 
Fig. 1 & $\lambda$ & 1.52 & 1.92 & 1.42 & 1.49 & 1.25 & 1.26 \\
       & $R$       & 0.97 & 0.79 & 1.05 & 0.87 & 1.25 & 1.10 \\
\hline
       & $\gamma$  & 0.76 & 0.80 & 0.95 & 1.00 & 0.97 & 1.02 \\ 
Fig. 2 & $\lambda$ & 1.53 & 1.58 & 0.41 & 0.59 & 0.21 & 0.33 \\
$(\tau=0.2)$ & $R$   & 0.67 & 0.94 & 0.60 & 1.34 & 0.46 & 1.32 \\
\hline
       & $\gamma$  & 0.49 & --- & 0.82 & --- & 0.93 & --- \\ 
Fig. 2 & $\lambda$ & 4.20 & --- & 1.27 & --- & 0.55 & --- \\
$(\tau=1/M)$ & $R$   & 0.60 & --- & 0.65 & --- & 0.65 & --- \\
\hline
\end{tabular}
\end{center}
\end{table}
\vspace{-2mm}

Our preliminary results are presented in Figs. 1 and 2 for
one-dimensional case only (first $3-$dimensional analysis of CAS is
provided in \cite{CAS1,CAS2}). The reference event for CAS was the
original cascade itself, which do not shows any BEC effect at all
(actually, the mixed-event method applied to MaxEnt gives in this
case identical result). To facilitate estimations of differences
between pictures presented in Figs. 1 and 2, we have listed in Table
1 parameters of simple exponential like parametrization of these
results. There are some features worth of noticing. Fig. 1
demonstrates that constant (i.e., independent on the details of the
hadronizing source given by event generators) weights lead to very
similar BEC pattern in both types of models. It is given entirely 
by the number of particles of the same charge in a given cell, 
which depends on $P$ (the bigger $P$ the more particles and bigger
$C_2(Q=0)$; smaller $P$ leads to increasing number of cells, which
results in decreasing $C_2(Q=0)$, as was already noticed in 
\cite{BSWW}). Only by
making $P$ depending on the details of hadronization process, as in
Fig. 2, we start to see differences between models. But even in this
case they are rather weak and $C_2$ depends  on the parameters of the
hadronic source only as much as they influence the number of
elementary cells and multiplicities in  them. Therefore the "size"
$R$ listed in Table 1 corresponds to the size of elementary cell
rather than to the size of the hadronizing source (in fact for
$\tau=0.2$ fm the real size of CAS source grows from $0.29$ fm for
$M=10$ GeV to $1.61$ fm for $M=100$ GeV and for $\tau=1/M$ from
$0.12$ fm to $0.62$ fm, respectively). There is noticeable difference
between CAS and MaxEnt cases for "slow" cascades ($\tau=0.2$ fm
\cite{CAS1}). For "fast" ones (with $\tau=1/M$, where particles are
produced much earlier and more uniformly in space-time) both the extensive MaxEnt and
nonextensive  CAS schemes lead to similar BEC. More detailed analysis
of this approach,  with its application also to the $3-$dimensional 
case will be presented elsewhere.

\section*{Acknowledgments}
One of the Authors (GW) would like to thank Tamas Cs\"org\H{o} and all
the Organizers of XXX-th ISMD for their kind hospitality and the
Hungarian Academy of Sciences for its financial support. The partial
support of Polish Committee for Scientific Research (grants 2P03B 011
18 and  621/E-78/SPUB/CERN/P-03/DZ4/99) is acknowledged.

\end{document}